\documentclass[aps,prl,twocolumn,showpacs,showkeys,superscriptaddress]{revtex4-1}

\usepackage{amsmath}
\usepackage[]{graphicx}
\usepackage{ifthen}

\input{MyMc.tex}

\newboolean{ShowCorrections}%without backslash
\setboolean{ShowCorrections}{false}%<-- for production, set this to false

\newcommand{\av}[2][]{\ensuremath{\left\langle#2\right\rangle_{#1}}\xspace}

\newcommand{\Sec}[1]{%
}
\newcommand{\Ack}{%
}

\begin{document}

\title{Experimental evidence of a $\varphi$ Josephson junction}

\author{H. Sickinger}
\affiliation{%
  Physikalisches Institut and Center for Collective Quantum Phenomena in LISA$^+$,
  Universit\"at T\"ubingen, Auf der Morgenstelle 14, D-72076 T\"ubingen, Germany
}

\author{A. Lipman}
\affiliation{%
  The Raymond and Beverly Sackler School of Physics and Astronomy,
  Tel Aviv University, Tel Aviv 69978, Israel
}

\author{M. Weides}
\affiliation{%
  Peter Gr\"unberg Institute and JARA-Fundamentals of Future Information Technology,
  Forschungszentrum J\"ulich GmbH, 52425 J\"ulich, Germany
}
\altaffiliation{%
  Current address:
  Physikalisches Institut, Karlsruher Institut f\"ur Technologie, 76131 Karlsruhe, Germany
}

\author{R. G. Mints}
\affiliation{%
  The Raymond and Beverly Sackler School of Physics and Astronomy,
  Tel Aviv University, Tel Aviv 69978, Israel
}

\author{H. Kohlstedt}
\affiliation{%
  Nanoelektronik, Technische Fakult\"at,
  Christian-Albrechts-Universit\"at zu Kiel,
  D-24143 Kiel, Germany
  %Universit\"at zu Kiel, Technische Fakult\"at,
  %Institut f\"ur Elektrotechnik und Informationstechnik, Nanoelektronik,
  %Kaiserstra{\ss}e 2, 24143 Kiel, Germany
}

\author{D. Koelle}
\author{R. Kleiner}
\author{E. Goldobin}
\affiliation{%
  Physikalisches Institut and Center for Collective Quantum Phenomena in LISA$^+$,
  Universit\"at T\"ubingen, Auf der Morgenstelle 14, D-72076 T\"ubingen, Germany
}

\date{\today}

\begin{abstract}
  We demonstrate experimentally the existence of Josephson junctions having a doubly degenerate ground state with an average Josephson phase $\psi=\pm\varphi$. The value of $\varphi$ can be chosen by design in the interval $0<\varphi<\pi$. The junctions used in our experiments are fabricated as 0-$\pi$ Josephson junctions of moderate normalized length with asymmetric 0 and $\pi$ regions. We show that (a) these $\varphi$ Josephson junctions have two critical currents, corresponding to the escape of the phase $\psi$ from $-\varphi$ and $+\varphi$ states; (b) the phase $\psi$ can be set to a particular state by tuning an external magnetic field or (c) by using a proper bias current sweep sequence. The experimental observations are in agreement with previous theoretical predictions.

\end{abstract}

\pacs{
  74.50.+r,   %Proximity effects, weak links, tunneling phenomena,
              %and Josephson effect
  85.25.Cp    %Josephson devices
%  05.45.-a    %Nonlinear dynamics and chaos
%  74.20.Rp    %Pairing symmetries (other than s-wave)
}

\keywords{$\varphi$ Josephson junction}

\maketitle

\Sec{Introduction}
\label{Sec:Intro}

%Intro what is 0 and pi JJs and why pi JJs are useful.
Josephson junctions (JJs) with a phase shift of $\pi$ in the ground state\cite{Bulaevskii:pi-loop} attracted a lot of interest in recent years\cite{Baselmans:1999:SNS-pi-JJ,Ryazanov:2001:SFS-PiJJ,Kontos:2002:SIFS-PiJJ,Weides:2006:SIFS-HiJcPiJJ,vanDam:2006:QuDot:SuperCurrRev,Gumann:2007:Geometric-pi-JJ}. In particular, these JJs can be used as on-chip phase batteries for biasing various classical\cite{Ortlepp:2006:RSFQ-0-pi} and quantum\cite{Feofanov:2010:SFS:pi-qubit} circuits, allowing for removing external bias lines and reducing decoherence.
Currently, it is possible to fabricate simultaneously both 0 and $\pi$ JJs using various technologies such as superconductor-ferromagnet heterostructures\cite{Ryazanov:2001:SFS-PiArray,Weides:2006:SIFS-0-pi,Frolov:2008:pi-arrays:Img,Weides:2007:JJ:TaylorBarrier,Pfeiffer:2008:SIFS-0-pi:HIZFS} or JJs based on d-wave superconductors\cite{VanHarlingen:1995:Review,Tsuei:Review,Smilde:ZigzagPRL,Guerlich:2009:LTSEM-zigzag}.

%??? 2 extra cites above

\begin{figure}[!tb]
  \centering\includegraphics{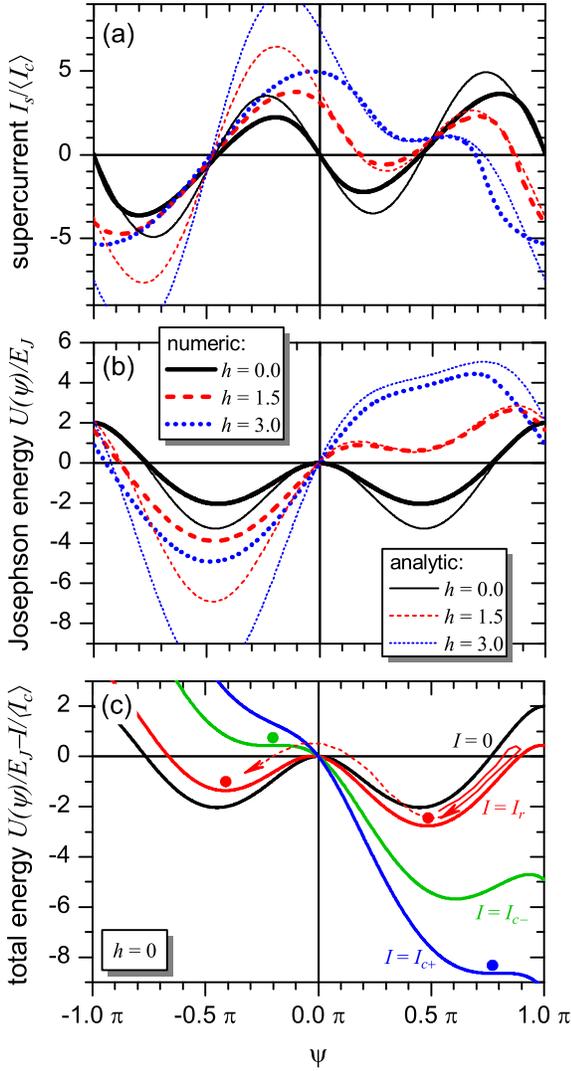}
  \caption{(Color online)
    (a) effective CPR $j_s(\psi)$ and (b) effective Josephson energy $U(\psi)$ calculated numerically (thick lines) in comparison to those given by approximate analytical formulas \eqref{Eq:CPR:sin+sin2+cos} and \eqref{Eq:U(psi)} (thin lines). For $h<0$ the $U(\psi)$ curves look mirror reflected with respect to the $U$-axis, \ie, $U(\psi,h)=U(-\psi,-h)$. (c) numerically calculated effective Josephson energy $U(\psi)$ at $h=0$ tilted by an applied bias current $I>0$ (energy supplied by the current source $-\Phi_0 I \psi/(2\pi)$). Several important situations are shown: ground state $I=0$ (same curve as in (b)), $I=I_r$ (retrapping), $I=I_{c-}$ (escape from $-\varphi$ well), and $I=I_{c+}$ (escape from $+\varphi$ well).
    \iffalse
    \EG{Guys, we have a problem here. In you CPR at $h=0$ the ratio of $I_{c+}/I_{c-}=4.59/3.27=1.40$In our experiment or direct simulations, see Fig.~\ref{Fig:IcH}, it is $I_{c+}/I_{c-}=561/327=1.71$! hmm??? Below I explain how did I get normalized length. The values of $j_{c,0}=62.9\units{A/cm^2}$ and $j_{c,\pi}=-47.9$ (note minus! But in Adi's simulation this should be plus) and $\ell=L/\lambda_J(\av{|j_c(x)|})=3.7$ (note normalization) are obtained from the fits. Then, from obtained $j_c$'s I calculated $\av{|j_c(x)|}=55.4\units{A/cm^2}$ and $\av{j_c(x)}=15\units{A/cm^2}$. Now, since $\ell$ is normalized using $\av{|j_c(x)|}$, but we want the length $l$ normalized to $\av{j_c(x)}$, I calculate\\
    $
      l= \ell \frac{\lambda_J(\av{|j_c(x)|})}{\lambda_J(\av{j_c(x)})}
      = \ell \sqrt{ \frac{\av{j_c(x)}}{\av{|j_c(x)|}} }
      = \ell \sqrt{ \frac{15.0}{55.4} }
      = 3.7\times 0.52
      = 1.925,
    $\\
    which makes 0.963 per segment.
    }
    \fi
  }
  \label{Fig:CPR&En}
\end{figure}

It would be remarkable to have a \emph{phase battery} providing an \emph{arbitrary} phase shift $\varphi$, rather than just 0 or $\pi$. The simplest idea is to combine 0 and $\pi$ JJs to obtain a $\varphi$ JJ. However, this is not as straightforward as it may seem. The balance between 0 and $\pi$ parts is complicated as shown in the pioneering work\cite{Bulaevskii:0-pi-LJJ}, where conditions for having a non-trivial $\varphi$-state were derived. Long artificial\cite{Buzdin:2003:phi-LJJ,Pugach:2010:CleanSFS:varphi-JJ} and natural\cite{Mints:1998:SelfGenFlux@AltJc,Ilichev:1999:InhomoPos2ndHarm,Mints:2002:SplinteredVortices@GB} arrays of $\ldots$-0-$\pi$-0-$\pi$-$\ldots$ JJs with short segments were analyzed in detail and suggested as systems, where a $\varphi$ JJ could be realized. More recently, only one period of such an array, \ie, one 0-$\pi$ JJ, was analyzed \emph{in an external magnetic field}\cite{Goldobin:2011:0-pi:H-tunable-CPR}. In these works, the Josephson phase $\phi(x)$ is considered as a sum of a constant (or slowly varying) phase $\psi$ and a deviation $|\xi(x)|\ll 1$ from the average phase. Then, for the average phase $\psi$ one obtains an effective current-phase relation (CPR) for the supercurrent\cite{Goldobin:2011:0-pi:H-tunable-CPR}
\begin{equation}
  I_s = \av{I_c}\left[ \sin(\psi) + \frac{\Gamma_0}{2} \sin(2\psi) + \Gamma_h h \cos(\psi)\right]
  , \label{Eq:CPR:sin+sin2+cos}
\end{equation}
where the averaged value of the critical current $\av{I_c}=\av{j_c}Lw$. The CPR \eqref{Eq:CPR:sin+sin2+cos} exactly corresponds to a $\varphi$ JJ at zero normalized magnetic field $h=0$, if $\Gamma_0<-1$, \cf Fig.\ref{Fig:CPR&En}(a). Here $L=L_0+L_\pi$ is the total length of JJ, while $L_0$ and $L_\pi$ are the lengths of 0 and $\pi$ parts, accordingly, $w$ is the width of JJ.

It is worth noting that the term ``$\varphi$ JJ'', introduced in Ref.~\onlinecite{Buzdin:2003:phi-LJJ}, refers to a JJ with a \emph{degenerate} ground state phase $\psi=\pm\varphi$. In the particular case of Eq.~\eqref{Eq:CPR:sin+sin2+cos} at $h=0$ one has $\varphi=\arccos(-1/2\Gamma_0)$. The coefficients $\Gamma_0$ and $\Gamma_h$ are defined as\cite{Lipman:varphiEx}
\begin{subequations}
  \begin{eqnarray}
    \Gamma_0 &=& -\frac{l_0^2 l_\pi^2}{3}\frac{(j_{c,0}-j_{c,\pi})^2}{(j_{c,0}l_0+j_{c,\pi} l_\pi)^2},\\
    \Gamma_h &=& \frac{l_0 l_\pi}{2}\,\frac{j_{c,0}-j_{c,\pi}}{j_{c,0} l_0+j_{c,\pi} l_\pi},
  \end{eqnarray}
  \label{Eq:Gammas}
\end{subequations}
where $l_0$, $l_\pi$ are the lengths normalized to the Josephson length $\lambda_J(\av{j_c})$ and $j_{c,0}$, $j_{c,\pi}$ are the critical current densities of 0 and $\pi$ parts, respectively. Here $\lambda_J$ is calculated using the average value of the critical current $\av{j_c}=(L_0 j_{c,0}+L_\pi j_{c,\pi})/(L_0+L_\pi)$.

The physics of $\varphi$ JJs with a CPR given by Eq.~\eqref{Eq:CPR:sin+sin2+cos} is quite unusual\cite{Goldobin:CPR:2ndHarm}. In particular, one should observe two critical currents\cite{Goldobin:CPR:2ndHarm,Goldobin:2011:0-pi:H-tunable-CPR} at $h=0$, corresponding to the escape of the phase from the left ($-\varphi$) or the right ($+\varphi$) well of the double-well Josephson energy potential
\begin{equation}
  U(\psi) = \av{E_J}\left[
    1-\cos(\psi) + \Gamma_h h \sin(\psi) + \frac{\Gamma_0}{2} \sin^2 (\psi)
  \right]
  , \label{Eq:U(psi)}
\end{equation}
where $\av{E_J}=\Phi_0 \av{I_c}/(2\pi)$. The critical currents are different because the maximum slope (maximum supercurrent in Fig.~\ref{Fig:CPR&En}(a)) on the rhs (positive bias) of the $-\varphi$ well is smaller than the maximum slope (maximum supercurrent in Fig.~\ref{Fig:CPR&En}(a)) on the rhs of the $+\varphi$ well, see Fig.~\ref{Fig:CPR&En}(b).

\begin{figure}[!tb]
\centering
  \includegraphics{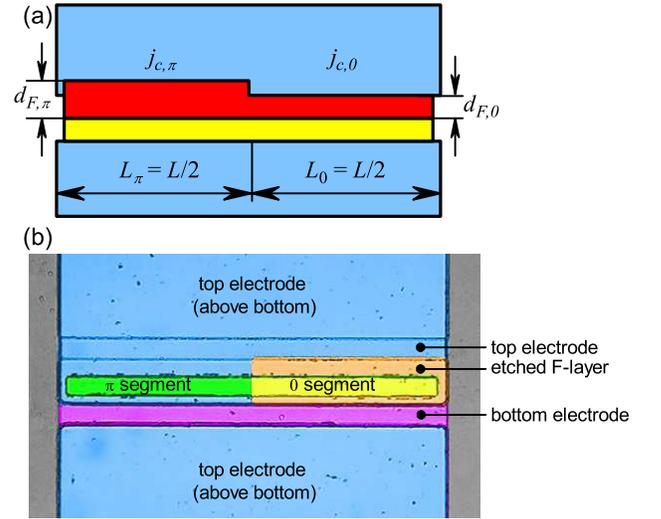}
  \caption{(Color online)
    (a) Sketch (cross section) of the investigated SIFS 0-$\pi$ JJ with a step in the F-layer thickness.
    (b) Optical image (top view, colored manually) of an investigated sample having overlap geometry. The junction area (0-segment and $\pi$-segment) is $200 \times 10 \units{\mu m^2}$. The ``etched F-layer'' area shows where the thickness of the F-layer was reduced from $d_{F,\pi}$ to $d_{F,0}$ to produce a 0-$\pi$ JJ.
  }
  \label{Fig:Sketch}
\end{figure}

In this letter we present experimental evidences of a $\varphi$ JJ made of one 0 and one $\pi$ segment, see Fig.~\ref{Fig:Sketch}(a).

\begin{figure}[!tb]
  \includegraphics{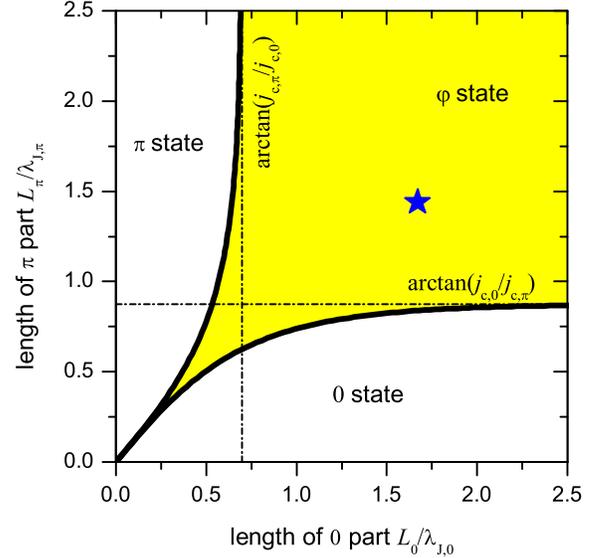}
  \caption{(Color online)
    Domain of existence of $\varphi$ state. The $\star$ shows the position of the investigated JJ at $T=2.35\units{K}$.
  }
  \label{Fig:domain}
\end{figure}

\Sec{Experiment}

The samples were fabricated as Nb$|$Al-Al$_2$O$_3$$|$Ni$_{0.6}$Cu$_{0.4}$$|$Nb heterostructures\cite{Weides:2006:SIFS-0-pi,Weides:2010:SIFS-jc1jc2:Ic(H)}. These superconductor-insulator-ferromagnet-superconductor (SIFS) JJs have an overlap geometry as shown in Fig.~\ref{Fig:Sketch}(b). Each junction consists of two parts, a conventional 0-segment and a $\pi$-segment. It is well known\cite{Kontos:2002:SIFS-PiJJ,Oboznov:2006:SFS-Ic(dF),Weides:2006:SIFS-HiJcPiJJ,Buzdin:2005:Review:SF} that the critical current in SFS or SIFS JJs strongly depends on the thickness $d_F$ of the F-layer and can become negative within some range of $d_F$ ($\pi$ junction). Therefore, to produce the 0 and the $\pi$ segments, the F-layer has different thicknesses $d_{F,0}$ and $d_{F,\pi}$, as shown in Fig.~\ref{Fig:Sketch}(a). To achieve this, the F-layer of thickness $d_{F,\pi}$, corresponding to $j_{c,\pi}\equiv j_c(d_{F,\pi})>0$ ($\pi$ JJ) was fabricated first. Then the area indicated in Fig.~\ref{Fig:Sketch}(b) was etched down to $d_{F,0}$, corresponding to $j_{c,0}\equiv j_c(d_{F,0})>0$ (0 JJ). Usually one obtains asymetric $j_c$ values\cite{Kemmler:2010:SIFS-0-pi:Ic(H)-asymm}, \ie $j_{c,0} \neq |j_{c,\pi}|$.

%??? Is citing Kemmler above appropriate?

We have studied 3 samples. One of them has very little $j_c$ asymmetry of about 5\%, and $L_0=L_\pi=25\units{\mu m}$, which corresponds to $L_0/\lambda_J(j_{c,0})\approx L_\pi/\lambda_J(j_{c,\pi})\approx0.37$. In this case even 5\% of asymmetry brings the sample out of the $\varphi$ domain, see Fig.~\ref{Fig:domain} for a qualitative picture. The other two JJs have $L_0=L_\pi=100\units{\mu m}$ and are deep inside the $\varphi$-domain in parameter space. Both samples show similar results. Here we present the results obtained on one of them. Its parameters are summarized in Tab.~\ref{Tab:samples} and its position within the $\varphi$ domain is indicated in Fig.~\ref{Fig:domain}. The values of $j_{c,0}$ and $j_{c,\pi}$ cannot be measured directly. In our case we have measured the $I_c(H)$ dependence (see below) and then simulated it numerically using $j_{c,0}$ and $j_{c,\pi}$ as fitting parameters. The best fitting was obtained for the values specified in Tab.~\ref{Tab:samples}.

\begin{table}[!tb]
  \caption{% $\av{|j_c(x)|}=55.4\units{A/cm^2}$, $\av{j_c(x)}=15.0\units{A/cm^2}$
    Junction parameters at $T=2.35\units{K}$. The values of $j_{c,0}$, $j_{c,\pi}$ and the normalized length $L/\lambda_J(\av{j_c(x)})$ are obtained from the fits.
    %First, from $j_c$'s and geometrical parameters of the films one can estimate $\ell=L/\lambda_J(\av{|j_c(x)|})\approx3.97$ and, taking the idle region into account, $\ell_\mathrm{eff}=L/\lambda_{J,\mathrm{eff}}(\av{|j_c(x)|})\approx3.13$. The value of $l=3.7$ obtained from the fit lays in between $\ell$ and $\ell_\mathrm{eff}$ and, in fact, the fits are not very sensitive to it.
    %Second, from $j_c$'s the values $\av{|j_c(x)|}=+55.4\units{A/cm^2}$ and $\av{j_c(x)}=+7.5\units{A/cm^2}$ can be calculated. In eqs. \eqref{Eq:Gammas} the lengths are normalized to $\lambda_J(\av{j_c(x)})=\sqrt{\av{|j_c(x)|}/\av{j_c(x)}}\lambda_J(\av{|j_c(x)|})=2.72\lambda_J(\av{|j_c(x)|})$.
  }
  \centering
  \begin{tabular}{lccc}
    \hline
    \hline
    parameter & $0$-part & $\pi$-part & whole JJ
    \\\hline
    physical length & $100 \units{\mu m}$ & $100\units{\mu m}$ & $200\units{\mu m}$
    \\
    %critical current density
    $j_{c,0}$, $j_{c,\pi}$, $\av{j_c(x)}$
    ($\units{A/cm^2}$)
    & $+62.9$ & $-47.9$ & $+7.5$
    \\
    normalized lengths %to $\lambda_J(\av{j_c(x)})$
    & 0.68 & 0.68 & 1.36\\
    \hline
    \hline
  \end{tabular}
  \label{Tab:samples}
\end{table}

\begin{figure}[!tb]
\includegraphics{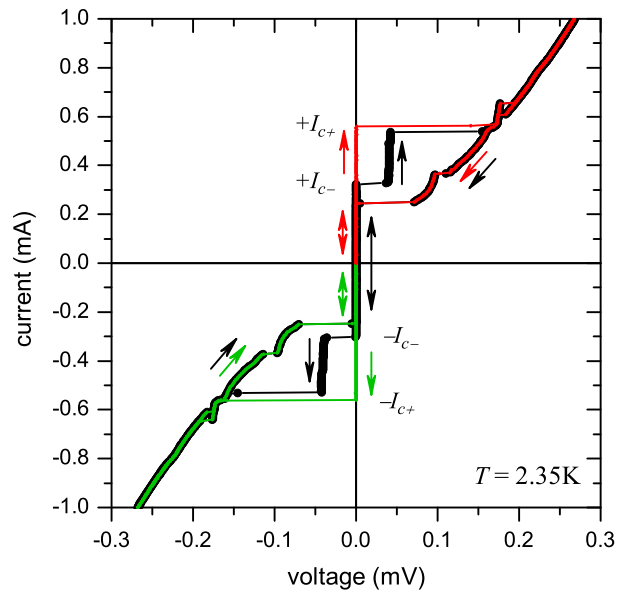}
  \caption{(Color online)
    Current-voltage characteristics showing lower $\pm I_{c-}$ and higher $\pm I_{c+}$ critical currents measured at $T\approx 2.35\units{K}$. At this temperature the behavior is deterministic:
    if one sweeps $I$ as $-I_\mathrm{max} \rightarrow +I_\mathrm{max} \rightarrow -I_\mathrm{max}$ one \emph{always} observes the critical currents $\pm I_{c-}$;
    if one sweeps $I$ as $+I_\mathrm{max}\rightarrow 0 \rightarrow +I_\mathrm{max}$, one \emph{always} observes $+I_{c+}$;
    finally, if one sweeps $I$ as $-I_\mathrm{max}\rightarrow0\rightarrow-I_\mathrm{max}$, one \emph{always} observes $-I_{c+}$.
  }
  \label{Fig:IVC}
\end{figure}

%IVC: theory vs. exp.
According to theoretical predictions\cite{Goldobin:CPR:2ndHarm,Goldobin:2011:0-pi:H-tunable-CPR} for a $\varphi$ JJ at zero magnetic field $H=0$ one expects two critical currents $\left|I_{c-}\right|<\left|I_{c+}\right|$ (for each bias direction), corresponding to the escape of the phase from $-\varphi$ and $+\varphi$ wells of $U(\psi)$ , respectively for bias current $I>0$ and vice versa for $I<0$. The current $I_{c+}$ is always observed. To observe $I_{c-}$ one has to have low damping so that retrapping in the $+\varphi$ well is avoided (for positive $I$). The $I$--$V$ characteristic (IVC) of the investigated 0-$\pi$ JJ at $H=0$ measured at $T\sim4.2\units{K}$ show only one critical current. Therefore the experiments were performed in a $^3$He cryostat at $T$ down to $300\units{mK}$ where the damping in SIFS JJs reduces drastically\cite{Pfeiffer:2008:SIFS-0-pi:HIZFS}. In the temperature range from $3.5\units{K}$ down to $300\units{mK}$ both $I_{c+}$ and $I_{c-}$ are clearly visible in the IVCs as shown in Fig.~\ref{Fig:IVC}.

Earlier\cite{Goldobin:CPR:2ndHarm} we proposed a technique that allows to choose which critical current one traces in the IVC, \ie from which well the phase escapes. The control is done by choosing a proper bias sweep sequence. For example, if the junction is returning from the positive voltage state, the potential $U(\psi)$ is tilted so that the phase slides to the right. When the tilt becomes small enough the phase will be trapped, presumably in the right $+\varphi$ well, \cf Fig.~\ref{Fig:CPR&En}(c). However, this natural assumption is not always true (see below). Then, if the phase is trapped at $+\varphi$, we can sweep the bias (a) in the positive direction and will observe escape from $+\varphi$ (to the right) at $I_{c+}$ or (b) in the negative direction to observe escape from $+\varphi$ (to the left) at $-I_{c-}$.

In experiment, at $T \lesssim 2.3\units{K}$, when the damping is very low, the currents $\pm I_{c+}$ and $\pm I_{c-}$ are traced in random order. Recording one IVC after the other, we were able to obtain IVCs with all 4 possible combinations: (a) $(-I_{c-},+I_{c-})$, (b) $(-I_{c-},+I_{c+})$, (c) $(-I_{c+},+I_{c-})$, (d) $(-I_{c+},+I_{c+})$. Choosing a specific sweep sequence as described above does not make the outcome ($I_{c-}$ or $I_{c+}$) predictable. We believe that in this temperature range the damping is so low that, upon returning from the positive voltage state, the phase does not simply stop in the $+\varphi$ well, but can also reflect from the barrier and find itself in a $-\varphi$ well, \cf Fig.~\ref{Fig:CPR&En}(c). The absence of determinism suggests that most probably we are dealing with a system exhibiting chaotic dynamics. This issue will be investigated elsewhere. Nevertheless, at $T\approx 2.35\units{K}$ we managed to achieve deterministic behavior as described above, see Fig.~\ref{Fig:IVC}.

%Namely, at zero field, if one sweeps $I$ from maximum $+I_\mathrm{max}$ to $I=0$ and back to $+I_\mathrm{max}$ one \emph{always} gets the higher critical current $+I_{c+}$ (for sweep $-I_\mathrm{max}\rightarrow 0 \rightarrow -I_\mathrm{max}$ one gets $-I_{c+}$). If one sweeps the current $I$ from $-I_\mathrm{max}$ to $+I_\mathrm{max}$ and back to $-I_\mathrm{max}$, one \emph{always} records the smaller critical current $\pm I_{c-}$.

%

Another fingerprint of a $\varphi$ JJ is its $I_c(H)$ dependence, which (a) should have the main cusp-like minima shifted off from $H=0$ point-symmetrically with respect to the origin, see Fig.~4 of Ref.~\onlinecite{Goldobin:2011:0-pi:H-tunable-CPR}; (b) should show up to four branches in total (for both sweep polarities) at low magnetic field\cite{Goldobin:2011:0-pi:H-tunable-CPR}. In essence, the latter feature results from the escape of the phase from two different energy minima in two different directions and is an extension of the two-critical-currents story to the case of non-zero magnetic field. Instead, the feature (a) alone cannot serve as a proof of a $\varphi$ JJ as it is a common feature of every asymmetric 0-$\pi$ JJ even if its ground state is 0 or $\pi$\cite{Goldobin:2011:0-pi:H-tunable-CPR}.

\begin{figure}[!tb]
  \includegraphics{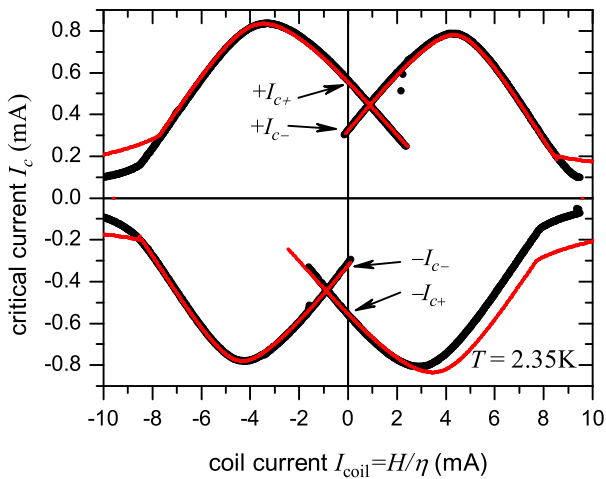}
  \caption{(Color online)
    Experimentally measured dependence $I_c(H)$ (black symbols) and numerically calculated curve (red/gray smaller symbols) that provides the best fit.
  }
  \label{Fig:IcH}
\end{figure}

The experimentally obtained $\pm I_c(H)$ dependence at $T=2.35\units{K}$ is shown in Fig.~\ref{Fig:IcH}. First, in the whole temperature range $0.3$--$4.2\units{K}$ we observe the main minima shifted point-symmetrically to a finite field, which is $H_\mathrm{min}=\eta \cdot I_\mathrm{coil}=\eta \cdot \pm879\units{\mu A}$. Second, at $T\lesssim 3.5\units{K}$ one observes the crossing of two branches and two critical currents for each bias current polarity. The left hand branches correspond to the escape of the phase out of the $+\varphi$-well, while the right branch corresponds to the escape from the $-\varphi$-well. Note, that at $T=4.2\units{K}$ the crossing of branches is not visible. One just observes a cusp-like minimum. In this case, both $I_{c\pm}$ cannot be seen together and the existence of two states, as should be present for a $\varphi$-JJ, cannot be proven.

%\EG{we may skip the following:}
To trace the intersection of the branches better we were applying a special value $H_\mathrm{reset}$ of the magnetic field during reset from the voltage $V>0$ state back to $V=0$ state, to ``prepare'' a specific state ($+\varphi$ or $-\varphi$) of the system. Then the field was set back to the ``current'' value $H$ and the bias was ramped up to trace $I_c(H)$. By doing this for different values of $H_\mathrm{reset}$, one is able to trace the intersection of the branches much better than without this technique, see Fig.~\ref{Fig:IcH}. Still, in experiment we were not able to trace the branches for positive and negative $I_c$ up to the point where they meet, as shown in Fig.~4(a) and (b) of Ref.~\onlinecite{Goldobin:2011:0-pi:H-tunable-CPR}, probably because of retrapping.

To extract some parameters of our JJ from the $I_c(H)$ curve, we performed numerical simulations, by solving the full sine-Gordon equation for a 0-$\pi$ JJ, using the normalized length $l$ and the critical currents $j_{c,0}$ and $j_{c,\pi}$ as fitting parameters. The simulations were performed using \textsc{StkJJ}\cite{StkJJ}. The objective was to obtain the best fit close to the origin, especially the point where the branches of $I_c(H)$ cross. The best fit is shown in Fig.~\ref{Fig:IcH} and was obtained for $l=3.7$ and the values $j_{c,0}$ and $j_{c,\pi}$ from Tab.~\ref{Tab:samples}. Measured data in Fig.~\ref{Fig:IcH} are shifted along $I_\mathrm{coil}$ axis by $530\units{\mu A}$ to compensate for average remanent magnetization of the F-layer. To obtain an almost perfect fit we have also assumed a difference in the constant remanent magnetizations in the 0 and in the $\pi$ parts of $\Delta\av{M}=\eta 1.5\units{mA}$, similar to earlier studies\cite{Kemmler:2010:SIFS-0-pi:Ic(H)-asymm,Weides:2010:SIFS-jc1jc2:Ic(H)}. One can also see that one of the experimental branches after the main maximum runs parallel to the simulated curve (the experimental $I_c(H)$ is not point symmetric). This shift stays the same even if we cycle the sample through $T_c$ of Nb, but changes if we cycle through the Curie temperature $T_C$ of the F-layer. We conclude that this is related to the non-uniform remanent magnetization of the F-layer.

The parameters obtained from the fit allow to see the location of our JJ on the $(l_0,l_\pi)$ plane, see Fig.~\ref{Fig:domain}. One sees that the JJ is not really short and lays quite deep in the $\varphi$ domain. In this region the analytic results\cite{Goldobin:2011:0-pi:H-tunable-CPR,Lipman:varphiEx} and, in particular, Eqs.~\eqref{Eq:CPR:sin+sin2+cos} and \eqref{Eq:U(psi)} are valid only qualitatively, \cf, Fig.~3 in Ref.~\onlinecite{Goldobin:2011:0-pi:H-tunable-CPR}. Therefore we calculated the CPR of our JJ numerically. We started with a 0-$\pi$ JJ with parameters $l_0$, $l_\pi$, $j_{c,0}$ and $j_{c,\pi}$ obtained from the fit, assumed some applied bias current $I$ and found \emph{all} static solutions $\phi(x)$ numerically. Then for each of these solutions we calculated $\psi\equiv\av{\phi(x)}$ and plotted all those $\psi$ on a $\psi(I)$ plot. By repeating this procedure for different $I$, we obtain an effective CPR $\psi(I)$ shown in Fig.~\ref{Fig:CPR&En}(a) together with the curve produced by the analytical formula \eqref{Eq:CPR:sin+sin2+cos}. One can see that the exact effective CPR calculated numerically and the approximate CPR calculated analytically are qualitatively similar. From the numerical CPR the value of $\varphi=0.41\pi$ for our particular $\varphi$ JJ is extracted.

%\EG{??? Make clear, why what we see is not a self-field effect. Simulations of self-field with StkJJ. Or arguments like no self-field is seen on 0 JJ.}

\Sec{Conclusions}
\label{Sec:Conclusions}

In conclusion, we have demonstrated experimentally the realization of a $\varphi$ Josephson junction based on a $0$-$\pi$ SIFS junction. We have observed experimentally a shift of the $I_c(H)$ minimum according to the effective CPR \eqref{Eq:CPR:sin+sin2+cos}, as predicted\cite{Goldobin:2011:0-pi:H-tunable-CPR}, as well as two critical currents $\pm I_{c\pm}$ close to zero magnetic field for each bias current polarity. We showed that one can choose between the $\pm\varphi$ states of the system using an externally applied magnetic field, which removes their degeneracy. We also showed that one can bring the system into one of the two states $\pm\varphi$ by properly tilting the potential using the bias current. Depending on the damping (temperature) one can achieve deterministic behavior (when damping is small enough to see the lower critical current, but large enough to trap the phase in a particular well) as well as random behavior at very low damping. The obtained JJ has $\varphi=0.41\pi$ at $I=0$).

\iffalse
\EG{It can provide the currents up to $I_{c-}=327\units{\mu A}$ ($\varphi$ drops to $0.22\pi$) or  $I_{c+}=561\units{\mu A}$ ($\varphi$ grows to $0.77\pi$) depending on initial state and polarity of the current}
\EG{Loading: The max loading has a difference between the case of applied current and connected inductance. In the 1st case $H=U(\psi)-I\psi$ (generalized washboard) and loading figures are as cited above. In the 2nd case $H=U(\psi)+\frac12 LI^2=U(\psi)+\frac12 (pre-factor)\psi^2$ and as current flow the $\psi$ decreases, which decreases current and so on. In this case for each inductance $L$ one can calculate the circulation current and find the critical value of inductance when the current will be zero.
This figures show loading capabilities of our $\varphi$ phase battery for different loading polarities.}
\fi

\Ack

We acknowledge financial support by the German Israeli Foundation (Grant No. G-967-126.14/2007) and DFG (via SFB/TRR-21, project A5 as well via project Go-1106/3). H.S. gratefully acknowledges support by the Evangelisches Studienwerk e.V. Villigst

\bibliography{SFS,SF,pi,LJJ,ratch,software}

\begin{thebibliography}{32}%
\makeatletter
\providecommand \@ifxundefined [1]{%
 \@ifx{#1\undefined}
}%
\providecommand \@ifnum [1]{%
 \ifnum #1\expandafter \@firstoftwo
 \else \expandafter \@secondoftwo
 \fi
}%
\providecommand \@ifx [1]{%
 \ifx #1\expandafter \@firstoftwo
 \else \expandafter \@secondoftwo
 \fi
}%
\providecommand \natexlab [1]{#1}%
\providecommand \enquote  [1]{``#1''}%
\providecommand \bibnamefont  [1]{#1}%
\providecommand \bibfnamefont [1]{#1}%
\providecommand \citenamefont [1]{#1}%
\providecommand \href@noop [0]{\@secondoftwo}%
\providecommand \href [0]{\begingroup \@sanitize@url \@href}%
\providecommand \@href[1]{\@@startlink{#1}\@@href}%
\providecommand \@@href[1]{\endgroup#1\@@endlink}%
\providecommand \@sanitize@url [0]{\catcode `\\12\catcode `\$12\catcode
  `\&12\catcode `\#12\catcode `\^12\catcode `\_12\catcode `\%12\relax}%
\providecommand \@@startlink[1]{}%
\providecommand \@@endlink[0]{}%
\providecommand \url  [0]{\begingroup\@sanitize@url \@url }%
\providecommand \@url [1]{\endgroup\@href {#1}{\urlprefix }}%
\providecommand \urlprefix  [0]{URL }%
\providecommand \Eprint [0]{\href }%
\providecommand \doibase [0]{http://dx.doi.org/}%
\providecommand \selectlanguage [0]{\@gobble}%
\providecommand \bibinfo  [0]{\@secondoftwo}%
\providecommand \bibfield  [0]{\@secondoftwo}%
\providecommand \translation [1]{[#1]}%
\providecommand \BibitemOpen [0]{}%
\providecommand \bibitemStop [0]{}%
\providecommand \bibitemNoStop [0]{.\EOS\space}%
\providecommand \EOS [0]{\spacefactor3000\relax}%
\providecommand \BibitemShut  [1]{\csname bibitem#1\endcsname}%
\let\auto@bib@innerbib\@empty
%</preamble>
\bibitem [{\citenamefont {Bulaevski\u{i}}\ \emph {et~al.}(1977)\citenamefont
  {Bulaevski\u{i}}, \citenamefont {Kuzi\u{i}},\ and\ \citenamefont
  {Sobyanin}}]{Bulaevskii:pi-loop}%
  \BibitemOpen
  \bibfield  {author} {\bibinfo {author} {\bibfnamefont {L.~N.}\ \bibnamefont
  {Bulaevski\u{i}}}, \bibinfo {author} {\bibfnamefont {V.~V.}\ \bibnamefont
  {Kuzi\u{i}}}, \ and\ \bibinfo {author} {\bibfnamefont {A.~A.}\ \bibnamefont
  {Sobyanin}},\ }\href@noop {} {\bibfield  {journal} {\bibinfo  {journal} {JETP
  Lett.}\ }\textbf {\bibinfo {volume} {25}},\ \bibinfo {pages} {290} (\bibinfo
  {year} {1977})},\ \bibinfo {note} {[Pis'ma Zh. Eksp. Teor. Fiz. 25, 314
  (1977)]}\BibitemShut {NoStop}%
\bibitem [{\citenamefont {Baselmans}\ \emph {et~al.}(1999)\citenamefont
  {Baselmans}, \citenamefont {Morpurgo}, \citenamefont {Wees},\ and\
  \citenamefont {Klapwijk}}]{Baselmans:1999:SNS-pi-JJ}%
  \BibitemOpen
  \bibfield  {author} {\bibinfo {author} {\bibfnamefont {J.~J.~A.}\
  \bibnamefont {Baselmans}}, \bibinfo {author} {\bibfnamefont {A.~F.}\
  \bibnamefont {Morpurgo}}, \bibinfo {author} {\bibfnamefont {B.~J.~V.}\
  \bibnamefont {Wees}}, \ and\ \bibinfo {author} {\bibfnamefont {T.~M.}\
  \bibnamefont {Klapwijk}},\ }\href {\doibase 10.1038/16204} {\bibfield
  {journal} {\bibinfo  {journal} {Nature (London)}\ }\textbf {\bibinfo {volume}
  {397}},\ \bibinfo {pages} {43} (\bibinfo {year} {1999})}\BibitemShut
  {NoStop}%
\bibitem [{\citenamefont {Ryazanov}\ \emph
  {et~al.}(2001{\natexlab{a}})\citenamefont {Ryazanov}, \citenamefont
  {Oboznov}, \citenamefont {Rusanov}, \citenamefont {Veretennikov},
  \citenamefont {Golubov},\ and\ \citenamefont
  {Aarts}}]{Ryazanov:2001:SFS-PiJJ}%
  \BibitemOpen
  \bibfield  {author} {\bibinfo {author} {\bibfnamefont {V.~V.}\ \bibnamefont
  {Ryazanov}}, \bibinfo {author} {\bibfnamefont {V.~A.}\ \bibnamefont
  {Oboznov}}, \bibinfo {author} {\bibfnamefont {A.~Y.}\ \bibnamefont
  {Rusanov}}, \bibinfo {author} {\bibfnamefont {A.~V.}\ \bibnamefont
  {Veretennikov}}, \bibinfo {author} {\bibfnamefont {A.~A.}\ \bibnamefont
  {Golubov}}, \ and\ \bibinfo {author} {\bibfnamefont {J.}~\bibnamefont
  {Aarts}},\ }\href@noop {} {\bibfield  {journal} {\bibinfo  {journal} {Phys.
  Rev. Lett.}\ }\textbf {\bibinfo {volume} {86}},\ \bibinfo {pages} {2427}
  (\bibinfo {year} {2001}{\natexlab{a}})}\BibitemShut {NoStop}%
\bibitem [{\citenamefont {Kontos}\ \emph {et~al.}(2002)\citenamefont {Kontos},
  \citenamefont {Aprili}, \citenamefont {Lesueur}, \citenamefont {Gen\^et},
  \citenamefont {Stephanidis},\ and\ \citenamefont
  {Boursier}}]{Kontos:2002:SIFS-PiJJ}%
  \BibitemOpen
  \bibfield  {author} {\bibinfo {author} {\bibfnamefont {T.}~\bibnamefont
  {Kontos}}, \bibinfo {author} {\bibfnamefont {M.}~\bibnamefont {Aprili}},
  \bibinfo {author} {\bibfnamefont {J.}~\bibnamefont {Lesueur}}, \bibinfo
  {author} {\bibfnamefont {F.}~\bibnamefont {Gen\^et}}, \bibinfo {author}
  {\bibfnamefont {B.}~\bibnamefont {Stephanidis}}, \ and\ \bibinfo {author}
  {\bibfnamefont {R.}~\bibnamefont {Boursier}},\ }\href@noop {} {\bibfield
  {journal} {\bibinfo  {journal} {Phys. Rev. Lett.}\ }\textbf {\bibinfo
  {volume} {89}},\ \bibinfo {pages} {137007} (\bibinfo {year}
  {2002})}\BibitemShut {NoStop}%
\bibitem [{\citenamefont {Weides}\ \emph
  {et~al.}(2006{\natexlab{a}})\citenamefont {Weides}, \citenamefont {Kemmler},
  \citenamefont {Goldobin}, \citenamefont {Koelle}, \citenamefont {Kleiner},
  \citenamefont {Kohlstedt},\ and\ \citenamefont
  {Buzdin}}]{Weides:2006:SIFS-HiJcPiJJ}%
  \BibitemOpen
  \bibfield  {author} {\bibinfo {author} {\bibfnamefont {M.}~\bibnamefont
  {Weides}}, \bibinfo {author} {\bibfnamefont {M.}~\bibnamefont {Kemmler}},
  \bibinfo {author} {\bibfnamefont {E.}~\bibnamefont {Goldobin}}, \bibinfo
  {author} {\bibfnamefont {D.}~\bibnamefont {Koelle}}, \bibinfo {author}
  {\bibfnamefont {R.}~\bibnamefont {Kleiner}}, \bibinfo {author} {\bibfnamefont
  {H.}~\bibnamefont {Kohlstedt}}, \ and\ \bibinfo {author} {\bibfnamefont
  {A.}~\bibnamefont {Buzdin}},\ }\href {\doibase 10.1063/1.2356104} {\bibfield
  {journal} {\bibinfo  {journal} {Appl. Phys. Lett.}\ }\textbf {\bibinfo
  {volume} {89}},\ \bibinfo {eid} {122511} (\bibinfo {year}
  {2006}{\natexlab{a}})},\ \Eprint {http://arxiv.org/abs/cond-mat/0604097}
  {cond-mat/0604097} \BibitemShut {NoStop}%
\bibitem [{\citenamefont {van Dam}\ \emph {et~al.}(2006)\citenamefont {van
  Dam}, \citenamefont {Nazarov}, \citenamefont {Bakkers}, \citenamefont
  {De~Franceschi},\ and\ \citenamefont
  {Kouwenhoven}}]{vanDam:2006:QuDot:SuperCurrRev}%
  \BibitemOpen
  \bibfield  {author} {\bibinfo {author} {\bibfnamefont {J.~A.}\ \bibnamefont
  {van Dam}}, \bibinfo {author} {\bibfnamefont {Y.~V.}\ \bibnamefont
  {Nazarov}}, \bibinfo {author} {\bibfnamefont {E.~P. A.~M.}\ \bibnamefont
  {Bakkers}}, \bibinfo {author} {\bibfnamefont {S.}~\bibnamefont
  {De~Franceschi}}, \ and\ \bibinfo {author} {\bibfnamefont {L.~P.}\
  \bibnamefont {Kouwenhoven}},\ }\href {\doibase 10.1038/nature05018}
  {\bibfield  {journal} {\bibinfo  {journal} {Nature (London)}\ }\textbf
  {\bibinfo {volume} {442}},\ \bibinfo {pages} {667} (\bibinfo {year}
  {2006})}\BibitemShut {NoStop}%
\bibitem [{\citenamefont {Gumann}\ \emph {et~al.}(2007)\citenamefont {Gumann},
  \citenamefont {Iniotakis},\ and\ \citenamefont
  {Schopohl}}]{Gumann:2007:Geometric-pi-JJ}%
  \BibitemOpen
  \bibfield  {author} {\bibinfo {author} {\bibfnamefont {A.}~\bibnamefont
  {Gumann}}, \bibinfo {author} {\bibfnamefont {C.}~\bibnamefont {Iniotakis}}, \
  and\ \bibinfo {author} {\bibfnamefont {N.}~\bibnamefont {Schopohl}},\ }\href
  {\doibase 10.1063/1.2801387} {\bibfield  {journal} {\bibinfo  {journal}
  {Appl. Phys. Lett.}\ }\textbf {\bibinfo {volume} {91}},\ \bibinfo {eid}
  {192502} (\bibinfo {year} {2007})}\BibitemShut {NoStop}%
\bibitem [{\citenamefont {Ortlepp}\ \emph {et~al.}(2006)\citenamefont
  {Ortlepp}, \citenamefont {Ariando}, \citenamefont {Mielke}, \citenamefont
  {Verwijs}, \citenamefont {Foo}, \citenamefont {Rogalla}, \citenamefont
  {Uhlmann},\ and\ \citenamefont {Hilgenkamp}}]{Ortlepp:2006:RSFQ-0-pi}%
  \BibitemOpen
  \bibfield  {author} {\bibinfo {author} {\bibfnamefont {T.}~\bibnamefont
  {Ortlepp}}, \bibinfo {author} {\bibnamefont {Ariando}}, \bibinfo {author}
  {\bibfnamefont {O.}~\bibnamefont {Mielke}}, \bibinfo {author} {\bibfnamefont
  {C.~J.~M.}\ \bibnamefont {Verwijs}}, \bibinfo {author} {\bibfnamefont
  {K.~F.~K.}\ \bibnamefont {Foo}}, \bibinfo {author} {\bibfnamefont
  {H.}~\bibnamefont {Rogalla}}, \bibinfo {author} {\bibfnamefont {F.~H.}\
  \bibnamefont {Uhlmann}}, \ and\ \bibinfo {author} {\bibfnamefont
  {H.}~\bibnamefont {Hilgenkamp}},\ }\href {\doibase 10.1126/science.1126041}
  {\bibfield  {journal} {\bibinfo  {journal} {Science}\ }\textbf {\bibinfo
  {volume} {312}},\ \bibinfo {pages} {1495} (\bibinfo {year}
  {2006})}\BibitemShut {NoStop}%
\bibitem [{\citenamefont {Feofanov}\ \emph {et~al.}(2010)\citenamefont
  {Feofanov}, \citenamefont {Oboznov}, \citenamefont {Bol'ginov}, \citenamefont
  {Lisenfeld}, \citenamefont {Poletto}, \citenamefont {Ryazanov}, \citenamefont
  {Rossolenko}, \citenamefont {Khabipov}, \citenamefont {Balashov},
  \citenamefont {Zorin}, \citenamefont {Dmitriev}, \citenamefont {Koshelets},\
  and\ \citenamefont {Ustinov}}]{Feofanov:2010:SFS:pi-qubit}%
  \BibitemOpen
  \bibfield  {author} {\bibinfo {author} {\bibfnamefont {A.~K.}\ \bibnamefont
  {Feofanov}}, \bibinfo {author} {\bibfnamefont {V.~A.}\ \bibnamefont
  {Oboznov}}, \bibinfo {author} {\bibfnamefont {V.~V.}\ \bibnamefont
  {Bol'ginov}}, \bibinfo {author} {\bibfnamefont {J.}~\bibnamefont
  {Lisenfeld}}, \bibinfo {author} {\bibfnamefont {S.}~\bibnamefont {Poletto}},
  \bibinfo {author} {\bibfnamefont {V.~V.}\ \bibnamefont {Ryazanov}}, \bibinfo
  {author} {\bibfnamefont {A.~N.}\ \bibnamefont {Rossolenko}}, \bibinfo
  {author} {\bibfnamefont {M.}~\bibnamefont {Khabipov}}, \bibinfo {author}
  {\bibfnamefont {D.}~\bibnamefont {Balashov}}, \bibinfo {author}
  {\bibfnamefont {A.~B.}\ \bibnamefont {Zorin}}, \bibinfo {author}
  {\bibfnamefont {P.~N.}\ \bibnamefont {Dmitriev}}, \bibinfo {author}
  {\bibfnamefont {V.~P.}\ \bibnamefont {Koshelets}}, \ and\ \bibinfo {author}
  {\bibfnamefont {A.~V.}\ \bibnamefont {Ustinov}},\ }\href {\doibase
  10.1038/nphys1700} {\bibfield  {journal} {\bibinfo  {journal} {Nat. Phys.}\
  }\textbf {\bibinfo {volume} {6}},\ \bibinfo {pages} {593} (\bibinfo {year}
  {2010})}\BibitemShut {NoStop}%
\bibitem [{\citenamefont {Ryazanov}\ \emph
  {et~al.}(2001{\natexlab{b}})\citenamefont {Ryazanov}, \citenamefont
  {Oboznov}, \citenamefont {Veretennikov},\ and\ \citenamefont
  {Rusanov}}]{Ryazanov:2001:SFS-PiArray}%
  \BibitemOpen
  \bibfield  {author} {\bibinfo {author} {\bibfnamefont {V.~V.}\ \bibnamefont
  {Ryazanov}}, \bibinfo {author} {\bibfnamefont {V.~A.}\ \bibnamefont
  {Oboznov}}, \bibinfo {author} {\bibfnamefont {A.~V.}\ \bibnamefont
  {Veretennikov}}, \ and\ \bibinfo {author} {\bibfnamefont {A.~Y.}\
  \bibnamefont {Rusanov}},\ }\href {\doibase 10.1103/PhysRevB.65.020501}
  {\bibfield  {journal} {\bibinfo  {journal} {Phys. Rev. B}\ }\textbf {\bibinfo
  {volume} {65}},\ \bibinfo {pages} {020501} (\bibinfo {year}
  {2001}{\natexlab{b}})}\BibitemShut {NoStop}%
\bibitem [{\citenamefont {Weides}\ \emph
  {et~al.}(2006{\natexlab{b}})\citenamefont {Weides}, \citenamefont {Kemmler},
  \citenamefont {Kohlstedt}, \citenamefont {Waser}, \citenamefont {Koelle},
  \citenamefont {Kleiner},\ and\ \citenamefont
  {Goldobin}}]{Weides:2006:SIFS-0-pi}%
  \BibitemOpen
  \bibfield  {author} {\bibinfo {author} {\bibfnamefont {M.}~\bibnamefont
  {Weides}}, \bibinfo {author} {\bibfnamefont {M.}~\bibnamefont {Kemmler}},
  \bibinfo {author} {\bibfnamefont {H.}~\bibnamefont {Kohlstedt}}, \bibinfo
  {author} {\bibfnamefont {R.}~\bibnamefont {Waser}}, \bibinfo {author}
  {\bibfnamefont {D.}~\bibnamefont {Koelle}}, \bibinfo {author} {\bibfnamefont
  {R.}~\bibnamefont {Kleiner}}, \ and\ \bibinfo {author} {\bibfnamefont
  {E.}~\bibnamefont {Goldobin}},\ }\href {\doibase
  10.1103/PhysRevLett.97.247001} {\bibfield  {journal} {\bibinfo  {journal}
  {Phys. Rev. Lett.}\ }\textbf {\bibinfo {volume} {97}},\ \bibinfo {eid}
  {247001} (\bibinfo {year} {2006}{\natexlab{b}})},\ \Eprint
  {http://arxiv.org/abs/cond-mat/0605656} {cond-mat/0605656} \BibitemShut
  {NoStop}%
\bibitem [{\citenamefont {Frolov}\ \emph {et~al.}(2008)\citenamefont {Frolov},
  \citenamefont {Stoutimore}, \citenamefont {Crane}, \citenamefont
  {Van~Harlingen}, \citenamefont {Oboznov}, \citenamefont {Ryazanov},
  \citenamefont {Ruosi}, \citenamefont {Granata},\ and\ \citenamefont
  {Russo}}]{Frolov:2008:pi-arrays:Img}%
  \BibitemOpen
  \bibfield  {author} {\bibinfo {author} {\bibfnamefont {S.~M.}\ \bibnamefont
  {Frolov}}, \bibinfo {author} {\bibfnamefont {M.~J.~A.}\ \bibnamefont
  {Stoutimore}}, \bibinfo {author} {\bibfnamefont {T.~A.}\ \bibnamefont
  {Crane}}, \bibinfo {author} {\bibfnamefont {D.~J.}\ \bibnamefont
  {Van~Harlingen}}, \bibinfo {author} {\bibfnamefont {V.~A.}\ \bibnamefont
  {Oboznov}}, \bibinfo {author} {\bibfnamefont {V.~V.}\ \bibnamefont
  {Ryazanov}}, \bibinfo {author} {\bibfnamefont {A.}~\bibnamefont {Ruosi}},
  \bibinfo {author} {\bibfnamefont {C.}~\bibnamefont {Granata}}, \ and\
  \bibinfo {author} {\bibfnamefont {M.}~\bibnamefont {Russo}},\ }\href
  {\doibase 10.1038/nphys780} {\bibfield  {journal} {\bibinfo  {journal} {Nat.
  Phys.}\ }\textbf {\bibinfo {volume} {4}},\ \bibinfo {pages} {32} (\bibinfo
  {year} {2008})}\BibitemShut {NoStop}%
\bibitem [{\citenamefont {Weides}\ \emph {et~al.}(2007)\citenamefont {Weides},
  \citenamefont {Schindler},\ and\ \citenamefont
  {Kohlstedt}}]{Weides:2007:JJ:TaylorBarrier}%
  \BibitemOpen
  \bibfield  {author} {\bibinfo {author} {\bibfnamefont {M.}~\bibnamefont
  {Weides}}, \bibinfo {author} {\bibfnamefont {C.}~\bibnamefont {Schindler}}, \
  and\ \bibinfo {author} {\bibfnamefont {H.}~\bibnamefont {Kohlstedt}},\ }\href
  {\doibase 10.1063/1.2655487} {\bibfield  {journal} {\bibinfo  {journal} {J.
  Appl. Phys.}\ }\textbf {\bibinfo {volume} {101}},\ \bibinfo {eid} {063902}
  (\bibinfo {year} {2007})}\BibitemShut {NoStop}%
\bibitem [{\citenamefont {Pfeiffer}\ \emph {et~al.}(2008)\citenamefont
  {Pfeiffer}, \citenamefont {Kemmler}, \citenamefont {Koelle}, \citenamefont
  {Kleiner}, \citenamefont {Goldobin}, \citenamefont {Weides}, \citenamefont
  {Feofanov}, \citenamefont {Lisenfeld},\ and\ \citenamefont
  {Ustinov}}]{Pfeiffer:2008:SIFS-0-pi:HIZFS}%
  \BibitemOpen
  \bibfield  {author} {\bibinfo {author} {\bibfnamefont {J.}~\bibnamefont
  {Pfeiffer}}, \bibinfo {author} {\bibfnamefont {M.}~\bibnamefont {Kemmler}},
  \bibinfo {author} {\bibfnamefont {D.}~\bibnamefont {Koelle}}, \bibinfo
  {author} {\bibfnamefont {R.}~\bibnamefont {Kleiner}}, \bibinfo {author}
  {\bibfnamefont {E.}~\bibnamefont {Goldobin}}, \bibinfo {author}
  {\bibfnamefont {M.}~\bibnamefont {Weides}}, \bibinfo {author} {\bibfnamefont
  {A.~K.}\ \bibnamefont {Feofanov}}, \bibinfo {author} {\bibfnamefont
  {J.}~\bibnamefont {Lisenfeld}}, \ and\ \bibinfo {author} {\bibfnamefont
  {A.~V.}\ \bibnamefont {Ustinov}},\ }\href {\doibase
  10.1103/PhysRevB.77.214506} {\bibfield  {journal} {\bibinfo  {journal} {Phys.
  Rev. B}\ }\textbf {\bibinfo {volume} {77}},\ \bibinfo {eid} {214506}
  (\bibinfo {year} {2008})},\ \Eprint {http://arxiv.org/abs/0801.3229}
  {0801.3229} \BibitemShut {NoStop}%
\bibitem [{\citenamefont {Van~Harlingen}(1995)}]{VanHarlingen:1995:Review}%
  \BibitemOpen
  \bibfield  {author} {\bibinfo {author} {\bibfnamefont {D.~J.}\ \bibnamefont
  {Van~Harlingen}},\ }\href {\doibase 10.1103/RevModPhys.67.515} {\bibfield
  {journal} {\bibinfo  {journal} {Rev. Mod. Phys.}\ }\textbf {\bibinfo {volume}
  {67}},\ \bibinfo {pages} {515} (\bibinfo {year} {1995})}\BibitemShut
  {NoStop}%
\bibitem [{\citenamefont {Tsuei}\ and\ \citenamefont
  {Kirtley}(2000)}]{Tsuei:Review}%
  \BibitemOpen
  \bibfield  {author} {\bibinfo {author} {\bibfnamefont {C.~C.}\ \bibnamefont
  {Tsuei}}\ and\ \bibinfo {author} {\bibfnamefont {J.~R.}\ \bibnamefont
  {Kirtley}},\ }\href {\doibase 10.1103/RevModPhys.72.969} {\bibfield
  {journal} {\bibinfo  {journal} {Rev. Mod. Phys.}\ }\textbf {\bibinfo {volume}
  {72}},\ \bibinfo {pages} {969} (\bibinfo {year} {2000})}\BibitemShut
  {NoStop}%
\bibitem [{\citenamefont {Smilde}\ \emph {et~al.}(2002)\citenamefont {Smilde},
  \citenamefont {Ariando}, \citenamefont {Blank}, \citenamefont {Gerritsma},
  \citenamefont {Hilgenkamp},\ and\ \citenamefont
  {Rogalla}}]{Smilde:ZigzagPRL}%
  \BibitemOpen
  \bibfield  {author} {\bibinfo {author} {\bibfnamefont {H.-J.~H.}\
  \bibnamefont {Smilde}}, \bibinfo {author} {\bibnamefont {Ariando}}, \bibinfo
  {author} {\bibfnamefont {D.~H.~A.}\ \bibnamefont {Blank}}, \bibinfo {author}
  {\bibfnamefont {G.~J.}\ \bibnamefont {Gerritsma}}, \bibinfo {author}
  {\bibfnamefont {H.}~\bibnamefont {Hilgenkamp}}, \ and\ \bibinfo {author}
  {\bibfnamefont {H.}~\bibnamefont {Rogalla}},\ }\href {\doibase
  10.1103/PhysRevLett.88.057004} {\bibfield  {journal} {\bibinfo  {journal}
  {Phys. Rev. Lett.}\ }\textbf {\bibinfo {volume} {88}},\ \bibinfo {pages}
  {057004} (\bibinfo {year} {2002})}\BibitemShut {NoStop}%
\bibitem [{\citenamefont {G\"{u}rlich}\ \emph {et~al.}(2009)\citenamefont
  {G\"{u}rlich}, \citenamefont {Goldobin}, \citenamefont {Straub},
  \citenamefont {Doenitz}, \citenamefont {Ariando}, \citenamefont {Smilde},
  \citenamefont {Hilgenkamp}, \citenamefont {Kleiner},\ and\ \citenamefont
  {Koelle}}]{Guerlich:2009:LTSEM-zigzag}%
  \BibitemOpen
  \bibfield  {author} {\bibinfo {author} {\bibfnamefont {C.}~\bibnamefont
  {G\"{u}rlich}}, \bibinfo {author} {\bibfnamefont {E.}~\bibnamefont
  {Goldobin}}, \bibinfo {author} {\bibfnamefont {R.}~\bibnamefont {Straub}},
  \bibinfo {author} {\bibfnamefont {D.}~\bibnamefont {Doenitz}}, \bibinfo
  {author} {\bibnamefont {Ariando}}, \bibinfo {author} {\bibfnamefont
  {H.-J.~H.}\ \bibnamefont {Smilde}}, \bibinfo {author} {\bibfnamefont
  {H.}~\bibnamefont {Hilgenkamp}}, \bibinfo {author} {\bibfnamefont
  {R.}~\bibnamefont {Kleiner}}, \ and\ \bibinfo {author} {\bibfnamefont
  {D.}~\bibnamefont {Koelle}},\ }\href {\doibase
  10.1103/PhysRevLett.103.067011} {\bibfield  {journal} {\bibinfo  {journal}
  {Phys. Rev. Lett.}\ }\textbf {\bibinfo {volume} {103}},\ \bibinfo {eid}
  {067011} (\bibinfo {year} {2009})}\BibitemShut {NoStop}%
\bibitem [{\citenamefont {Bulaevskii}\ \emph {et~al.}(1978)\citenamefont
  {Bulaevskii}, \citenamefont {Kuzii},\ and\ \citenamefont
  {Sobyanin}}]{Bulaevskii:0-pi-LJJ}%
  \BibitemOpen
  \bibfield  {author} {\bibinfo {author} {\bibfnamefont {L.~N.}\ \bibnamefont
  {Bulaevskii}}, \bibinfo {author} {\bibfnamefont {V.~V.}\ \bibnamefont
  {Kuzii}}, \ and\ \bibinfo {author} {\bibfnamefont {A.~A.}\ \bibnamefont
  {Sobyanin}},\ }\href@noop {} {\bibfield  {journal} {\bibinfo  {journal}
  {Solid State Commun.}\ }\textbf {\bibinfo {volume} {25}},\ \bibinfo {pages}
  {1053} (\bibinfo {year} {1978})}\BibitemShut {NoStop}%
\bibitem [{\citenamefont {Buzdin}\ and\ \citenamefont
  {Koshelev}(2003)}]{Buzdin:2003:phi-LJJ}%
  \BibitemOpen
  \bibfield  {author} {\bibinfo {author} {\bibfnamefont {A.}~\bibnamefont
  {Buzdin}}\ and\ \bibinfo {author} {\bibfnamefont {A.~E.}\ \bibnamefont
  {Koshelev}},\ }\href {\doibase 10.1103/PhysRevB.67.220504} {\bibfield
  {journal} {\bibinfo  {journal} {Phys. Rev. B}\ }\textbf {\bibinfo {volume}
  {67}},\ \bibinfo {eid} {220504(R)} (\bibinfo {year} {2003})},\ \Eprint
  {http://arxiv.org/abs/cond-mat/0305142} {cond-mat/0305142} \BibitemShut
  {NoStop}%
\bibitem [{\citenamefont {Pugach}\ \emph {et~al.}(2010)\citenamefont {Pugach},
  \citenamefont {Goldobin}, \citenamefont {Kleiner},\ and\ \citenamefont
  {Koelle}}]{Pugach:2010:CleanSFS:varphi-JJ}%
  \BibitemOpen
  \bibfield  {author} {\bibinfo {author} {\bibfnamefont {N.~G.}\ \bibnamefont
  {Pugach}}, \bibinfo {author} {\bibfnamefont {E.}~\bibnamefont {Goldobin}},
  \bibinfo {author} {\bibfnamefont {R.}~\bibnamefont {Kleiner}}, \ and\
  \bibinfo {author} {\bibfnamefont {D.}~\bibnamefont {Koelle}},\ }\href
  {\doibase 10.1103/PhysRevB.81.104513} {\bibfield  {journal} {\bibinfo
  {journal} {Phys. Rev. B}\ }\textbf {\bibinfo {volume} {81}},\ \bibinfo
  {pages} {104513} (\bibinfo {year} {2010})}\BibitemShut {NoStop}%
\bibitem [{\citenamefont {Mints}(1998)}]{Mints:1998:SelfGenFlux@AltJc}%
  \BibitemOpen
  \bibfield  {author} {\bibinfo {author} {\bibfnamefont {R.~G.}\ \bibnamefont
  {Mints}},\ }\href {\doibase 10.1103/PhysRevB.57.R3221} {\bibfield  {journal}
  {\bibinfo  {journal} {Phys. Rev. B}\ }\textbf {\bibinfo {volume} {57}},\
  \bibinfo {pages} {R3221} (\bibinfo {year} {1998})}\BibitemShut {NoStop}%
\bibitem [{\citenamefont {Il'ichev}\ \emph {et~al.}(1999)\citenamefont
  {Il'ichev}, \citenamefont {Zakosarenko}, \citenamefont {IJsselsteijn},
  \citenamefont {Hoenig}, \citenamefont {Meyer}, \citenamefont {Fistul},\ and\
  \citenamefont {M\"uller}}]{Ilichev:1999:InhomoPos2ndHarm}%
  \BibitemOpen
  \bibfield  {author} {\bibinfo {author} {\bibfnamefont {E.}~\bibnamefont
  {Il'ichev}}, \bibinfo {author} {\bibfnamefont {V.}~\bibnamefont
  {Zakosarenko}}, \bibinfo {author} {\bibfnamefont {R.~P.~J.}\ \bibnamefont
  {IJsselsteijn}}, \bibinfo {author} {\bibfnamefont {H.~E.}\ \bibnamefont
  {Hoenig}}, \bibinfo {author} {\bibfnamefont {H.-G.}\ \bibnamefont {Meyer}},
  \bibinfo {author} {\bibfnamefont {M.~V.}\ \bibnamefont {Fistul}}, \ and\
  \bibinfo {author} {\bibfnamefont {P.}~\bibnamefont {M\"uller}},\ }\href
  {\doibase 10.1103/PhysRevB.59.11502} {\bibfield  {journal} {\bibinfo
  {journal} {Phys. Rev. B}\ }\textbf {\bibinfo {volume} {59}},\ \bibinfo
  {pages} {11502} (\bibinfo {year} {1999})}\BibitemShut {NoStop}%
\bibitem [{\citenamefont {Mints}\ \emph {et~al.}(2002)\citenamefont {Mints},
  \citenamefont {Papiashvili}, \citenamefont {Kirtley}, \citenamefont
  {Hilgenkamp}, \citenamefont {Hammerl},\ and\ \citenamefont
  {Mannhart}}]{Mints:2002:SplinteredVortices@GB}%
  \BibitemOpen
  \bibfield  {author} {\bibinfo {author} {\bibfnamefont {R.~G.}\ \bibnamefont
  {Mints}}, \bibinfo {author} {\bibfnamefont {I.}~\bibnamefont {Papiashvili}},
  \bibinfo {author} {\bibfnamefont {J.~R.}\ \bibnamefont {Kirtley}}, \bibinfo
  {author} {\bibfnamefont {H.}~\bibnamefont {Hilgenkamp}}, \bibinfo {author}
  {\bibfnamefont {G.}~\bibnamefont {Hammerl}}, \ and\ \bibinfo {author}
  {\bibfnamefont {J.}~\bibnamefont {Mannhart}},\ }\href {\doibase
  10.1103/PhysRevLett.89.067004} {\bibfield  {journal} {\bibinfo  {journal}
  {Phys. Rev. Lett.}\ }\textbf {\bibinfo {volume} {89}},\ \bibinfo {pages}
  {067004} (\bibinfo {year} {2002})}\BibitemShut {NoStop}%
\bibitem [{\citenamefont {Goldobin}\ \emph {et~al.}(2011)\citenamefont
  {Goldobin}, \citenamefont {Koelle}, \citenamefont {Kleiner},\ and\
  \citenamefont {Mints}}]{Goldobin:2011:0-pi:H-tunable-CPR}%
  \BibitemOpen
  \bibfield  {author} {\bibinfo {author} {\bibfnamefont {E.}~\bibnamefont
  {Goldobin}}, \bibinfo {author} {\bibfnamefont {D.}~\bibnamefont {Koelle}},
  \bibinfo {author} {\bibfnamefont {R.}~\bibnamefont {Kleiner}}, \ and\
  \bibinfo {author} {\bibfnamefont {R.~G.}\ \bibnamefont {Mints}},\ }\href
  {\doibase 10.1103/PhysRevLett.107.227001} {\bibfield  {journal} {\bibinfo
  {journal} {Phys. Rev. Lett.}\ }\textbf {\bibinfo {volume} {107}},\ \bibinfo
  {pages} {227001} (\bibinfo {year} {2011})},\ \Eprint
  {http://arxiv.org/abs/1110.2326} {1110.2326} \BibitemShut {NoStop}%
\bibitem [{\citenamefont {Lipman}\ \emph {et~al.}(2012)\citenamefont {Lipman},
  \citenamefont {Goldobin}, \citenamefont {Koelle}, \citenamefont {Kleiner},\
  and\ \citenamefont {Mints}}]{Lipman:varphiEx}%
  \BibitemOpen
  \bibfield  {author} {\bibinfo {author} {\bibfnamefont {A.}~\bibnamefont
  {Lipman}}, \bibinfo {author} {\bibfnamefont {E.}~\bibnamefont {Goldobin}},
  \bibinfo {author} {\bibfnamefont {D.}~\bibnamefont {Koelle}}, \bibinfo
  {author} {\bibfnamefont {R.}~\bibnamefont {Kleiner}}, \ and\ \bibinfo
  {author} {\bibfnamefont {R.}~\bibnamefont {Mints}},\ }\href@noop {} {\enquote
  {\bibinfo {title} {Josephson junction with a magnetic-field tunable
  current-phase relation},}\ } (\bibinfo {year} {2012}),\ \bibinfo {note}
  {unpublished}\BibitemShut {NoStop}%
\bibitem [{\citenamefont {Goldobin}\ \emph {et~al.}(2007)\citenamefont
  {Goldobin}, \citenamefont {Koelle}, \citenamefont {Kleiner},\ and\
  \citenamefont {Buzdin}}]{Goldobin:CPR:2ndHarm}%
  \BibitemOpen
  \bibfield  {author} {\bibinfo {author} {\bibfnamefont {E.}~\bibnamefont
  {Goldobin}}, \bibinfo {author} {\bibfnamefont {D.}~\bibnamefont {Koelle}},
  \bibinfo {author} {\bibfnamefont {R.}~\bibnamefont {Kleiner}}, \ and\
  \bibinfo {author} {\bibfnamefont {A.}~\bibnamefont {Buzdin}},\ }\href
  {\doibase 10.1103/PhysRevB.76.224523} {\bibfield  {journal} {\bibinfo
  {journal} {Phys. Rev. B}\ }\textbf {\bibinfo {volume} {76}},\ \bibinfo {eid}
  {224523} (\bibinfo {year} {2007})},\ \Eprint {http://arxiv.org/abs/0708.2624}
  {0708.2624} \BibitemShut {NoStop}%
\bibitem [{\citenamefont {Weides}\ \emph {et~al.}(2010)\citenamefont {Weides},
  \citenamefont {Peralagu}, \citenamefont {Kohlstedt}, \citenamefont
  {Pfeiffer}, \citenamefont {Kemmler}, \citenamefont {G\"urlich}, \citenamefont
  {Goldobin}, \citenamefont {Koelle},\ and\ \citenamefont
  {Kleiner}}]{Weides:2010:SIFS-jc1jc2:Ic(H)}%
  \BibitemOpen
  \bibfield  {author} {\bibinfo {author} {\bibfnamefont {M.}~\bibnamefont
  {Weides}}, \bibinfo {author} {\bibfnamefont {U.}~\bibnamefont {Peralagu}},
  \bibinfo {author} {\bibfnamefont {H.}~\bibnamefont {Kohlstedt}}, \bibinfo
  {author} {\bibfnamefont {J.}~\bibnamefont {Pfeiffer}}, \bibinfo {author}
  {\bibfnamefont {M.}~\bibnamefont {Kemmler}}, \bibinfo {author} {\bibfnamefont
  {C.}~\bibnamefont {G\"urlich}}, \bibinfo {author} {\bibfnamefont
  {E.}~\bibnamefont {Goldobin}}, \bibinfo {author} {\bibfnamefont
  {D.}~\bibnamefont {Koelle}}, \ and\ \bibinfo {author} {\bibfnamefont
  {R.}~\bibnamefont {Kleiner}},\ }\href {\doibase
  10.1088/0953-2048/23/9/095007} {\bibfield  {journal} {\bibinfo  {journal}
  {Supercond. Sci. Technol.}\ }\textbf {\bibinfo {volume} {23}},\ \bibinfo
  {pages} {095007} (\bibinfo {year} {2010})}\BibitemShut {NoStop}%
\bibitem [{\citenamefont {Oboznov}\ \emph {et~al.}(2006)\citenamefont
  {Oboznov}, \citenamefont {Bol'ginov}, \citenamefont {Feofanov}, \citenamefont
  {Ryazanov},\ and\ \citenamefont {Buzdin}}]{Oboznov:2006:SFS-Ic(dF)}%
  \BibitemOpen
  \bibfield  {author} {\bibinfo {author} {\bibfnamefont {V.~A.}\ \bibnamefont
  {Oboznov}}, \bibinfo {author} {\bibfnamefont {V.~V.}\ \bibnamefont
  {Bol'ginov}}, \bibinfo {author} {\bibfnamefont {A.~K.}\ \bibnamefont
  {Feofanov}}, \bibinfo {author} {\bibfnamefont {V.~V.}\ \bibnamefont
  {Ryazanov}}, \ and\ \bibinfo {author} {\bibfnamefont {A.~I.}\ \bibnamefont
  {Buzdin}},\ }\href {\doibase 10.1103/PhysRevLett.96.197003} {\bibfield
  {journal} {\bibinfo  {journal} {Phys. Rev. Lett.}\ }\textbf {\bibinfo
  {volume} {96}},\ \bibinfo {eid} {197003} (\bibinfo {year}
  {2006})}\BibitemShut {NoStop}%
\bibitem [{\citenamefont {Buzdin}(2005)}]{Buzdin:2005:Review:SF}%
  \BibitemOpen
  \bibfield  {author} {\bibinfo {author} {\bibfnamefont {A.~I.}\ \bibnamefont
  {Buzdin}},\ }\href@noop {} {\bibfield  {journal} {\bibinfo  {journal} {Rev.
  Mod. Phys.}\ }\textbf {\bibinfo {volume} {77}},\ \bibinfo {eid} {935}
  (\bibinfo {year} {2005})}\BibitemShut {NoStop}%
\bibitem [{\citenamefont {Kemmler}\ \emph {et~al.}(2010)\citenamefont
  {Kemmler}, \citenamefont {Weides}, \citenamefont {Weiler}, \citenamefont
  {Opel}, \citenamefont {Goennenwein}, \citenamefont {Vasenko}, \citenamefont
  {Golubov}, \citenamefont {Kohlstedt}, \citenamefont {Koelle}, \citenamefont
  {Kleiner},\ and\ \citenamefont
  {Goldobin}}]{Kemmler:2010:SIFS-0-pi:Ic(H)-asymm}%
  \BibitemOpen
  \bibfield  {author} {\bibinfo {author} {\bibfnamefont {M.}~\bibnamefont
  {Kemmler}}, \bibinfo {author} {\bibfnamefont {M.}~\bibnamefont {Weides}},
  \bibinfo {author} {\bibfnamefont {M.}~\bibnamefont {Weiler}}, \bibinfo
  {author} {\bibfnamefont {M.}~\bibnamefont {Opel}}, \bibinfo {author}
  {\bibfnamefont {S.~T.~B.}\ \bibnamefont {Goennenwein}}, \bibinfo {author}
  {\bibfnamefont {A.~S.}\ \bibnamefont {Vasenko}}, \bibinfo {author}
  {\bibfnamefont {A.~A.}\ \bibnamefont {Golubov}}, \bibinfo {author}
  {\bibfnamefont {H.}~\bibnamefont {Kohlstedt}}, \bibinfo {author}
  {\bibfnamefont {D.}~\bibnamefont {Koelle}}, \bibinfo {author} {\bibfnamefont
  {R.}~\bibnamefont {Kleiner}}, \ and\ \bibinfo {author} {\bibfnamefont
  {E.}~\bibnamefont {Goldobin}},\ }\href {\doibase 10.1103/PhysRevB.81.054522}
  {\bibfield  {journal} {\bibinfo  {journal} {Phys. Rev. B}\ }\textbf {\bibinfo
  {volume} {81}},\ \bibinfo {pages} {054522} (\bibinfo {year}
  {2010})}\BibitemShut {NoStop}%
\bibitem [{\citenamefont {Goldobin}(2011)}]{StkJJ}%
  \BibitemOpen
  \bibfield  {author} {\bibinfo {author} {\bibfnamefont {E.}~\bibnamefont
  {Goldobin}},\ }\href@noop {} {\enquote {\bibinfo {title} {\textsc{StkJJ --
  User's Reference}},}\ }\bibinfo {howpublished}
  {\url{http://www.geocities.com/SiliconValley/Heights/7318/StkJJ.htm}}
  (\bibinfo {year} {2011})\BibitemShut {NoStop}%
\end{thebibliography}%

\end{document}